\documentclass[journal]{IEEEtran}
\usepackage{amsbsy}
\usepackage{graphics}
\usepackage{graphicx}
\usepackage{amsfonts}
\usepackage{amsmath}
\usepackage{amssymb}
\usepackage{fancyhdr}
\usepackage{color}

\newtheorem{theorem}{Theorem}

\renewcommand{\baselinestretch}{2}
\onecolumn

\begin{document}
\large
\title{A Compressed PCA Subspace Method for Anomaly Detection
	in High-Dimensional Data}
\author{Qi Ding and Eric D. Kolaczyk\thanks{Qi Ding and Eric D. Kolaczyk
are with the Department of Mathematics and Statistics, Boston
University, Boston, MA 02215 USA (email: qiding@bu.edu;
kolaczyk@bu.edu). The authors thank Debashis Paul for a number of helpful conversations.  This work was supported in part by NSF grants CCF-0325701 and CNS-0905565 and by ONR award N000140910654.}, {\it Senior Member, IEEE}}

\maketitle

\begin{abstract}
Random projection is widely used as a method of dimension
reduction.  In recent years, its combination with standard
techniques of regression and classification has been explored.
Here we examine its use for anomaly detection in high-dimensional settings,  in conjunction
with principal component analysis (PCA) and corresponding subspace detection methods. 
We assume a so-called spiked covariance model for the underlying data generation process
and a Gaussian random projection.  We adopt a hypothesis testing
perspective of the anomaly detection problem, with the test statistic defined to be
the magnitude of the residuals of a PCA analysis.  
Under the null hypothesis of no anomaly, we characterize the relative accuracy with which the mean and variance
of the test statistic from compressed data approximate those of the corresponding test
statistic from uncompressed data.  Furthermore, under a suitable alternative hypothesis, we provide expressions that
allow for a comparison of statistical power for detection.  Finally, whereas these results correspond to the ideal setting
in which the data covariance is known, we show that it is possible to obtain the same order of accuracy when the 
covariance of the compressed measurements is estimated using a sample covariance, as long as the number of measurements is of the same order of 
magnitude as the reduced dimensionality.
\end{abstract}

\section{Introduction}
\label{sec:intro}

Principal component analysis (PCA) is a classical tool for dimension reduction that remains at
the heart of many modern techniques in multivariate statistics and data mining.  Among the multitude of uses that have
been found for it, PCA often plays a central role in methods for systems monitoring and anomaly
detection.  A prototypical example of this is the method of Jackson and Mudholkar~\cite{jackson.1979}, the
so-called PCA subspace projection method.  In their approach,
PCA is used to extract the primary trends and patterns in data and the magnitude of the
residuals (i.e., the norm of the projection of the data into the residual subspace)
is then monitored for departures,  with principles from hypothesis testing being used to set detection thresholds.
This method has seen widespread usage in industrial systems control (e.g.\cite{crb,nomikos.macgregor,qin2003statistical}).
More recently, it is also being used in the analysis of financial data (e.g.~\cite{harvey.etal,laloux.etal,lessard}) and
of Internet traffic data (e.g. \cite{lakhina.etal.2004,lakhina.etal.2005}).

In this paper, we propose a methodology in which PCA subspace projection is applied to data
that have first undergone random projection.
Two key observations motivate this proposal.  First, as is well-known, the computational complexity of
PCA, when computed using the standard approach based on the singular value decomposition,
scales like $O(l^3+l^2 n)$, where $l$ is the dimensionality of the data
and $n$ is the sample size.   Thus use of the PCA subspace method
is increasingly less feasible with the ever-increasing size nd dimensions 
of modern data sets.  Second, concerns
regarding data confidentiality, whether for proprietary reasons or reasons of privacy, are more and more
driving a need for statistical methods to accommodate.
The first of these problems is something a number of authors have sought to address in recent years
(e.g., \cite{zht,johnstone.lu,shen2008sparse,journee2010generalized}), 
while the second, of course, does not pertain to PCA-based methods alone.  Our
proposal to incorporate random projection into the PCA subspace method is made with both issues in mind,
in that the original data are transformed to a random coordinate space of reduced dimension prior to being
processed.

The key application motivating our problem is that of monitoring Internet traffic data.  Previous use of
PCA subspace methods for traffic monitoring~\cite{lakhina.etal.2004,lakhina.etal.2005}
has been largely restricted to the level of traffic traces aggregated
over broad metropolitan regions (e.g., New York, Chicago, Los Angeles) for a network covering
an entire country or continent (e.g., the United States, Europe, etc.).  This level of aggregation is useful for monitoring
coarse-scale usage patterns and high-level quality-of-service obligations.  However, much of the current
interest in the analysis of Internet traffic data revolves around the much finer scale of individual
users.  Data of this sort can be determined up to the (apparent) identity of individual computing devices, i.e., so-called
IP addresses.  But there are as many as $2^{32}$ such IP address, making the monitoring of traffic at this level
a task guaranteed to involve massive amounts of data of very high dimension.  
Furthermore, it is typically necessary to anonymize data
of this sort, and often it is not possible for anyone outside of the auspices of a particular Internet service provider
to work with such data in its original form.  The standard technique used when data of this sort are actually shared
is to aggregate the IP addresses in a manner similar to the coarsening of geo-coding (e.g., giving only information on
a town of residence, rather than a street address).  Our proposed methodology can be viewed as a stylized
prototype, establishing proof-of-concept for the use of PCA subspace projection methods on data like IP-level
Internet traffic in a way that is both computationally feasible and respects concerns for data
confidentiality.  

Going back to the famous Johnson-Lindenstrauss lemma~\cite{johnson.lindenstrauss}, it is now
well-known that an appropriately defined random projection will effectively preserve length of data vectors as well
as distance between vectors.  This fact lies at the heart of an explosion in recent years of new theory and methods in
statistics, machine learning, and signal processing.  These include~\cite{dasgupta,fradkin.madigan,bingham.Mannula}.
See, for example, the review~\cite{vempala}.  Many of these
methods go by names emphasizing the compression inherent in the random projection, such as
`compressed sensing' or `compressive sampling'.  In this spirit, we call our own method
{\em compressed PCA subspace projection}.
The primary contribution of our work is to show that, under certain sparseness conditions on the covariance
structure of the original data, the use of Gaussian random projection followed by projection into the PCA residual subspace
yields a test statistic $Q^*$ whose distributional behavior is comparable to that of the statistic $Q$ that would have
been obtained from PCA subspace projection on the original data.  And furthermore that, up to higher order terms, there is
no loss in accuracy if an estimated covariance matrix is used, rather than the true (unknown) covariance, as long as the 
sample size for estimating the covariance is of the same order of magnitude as dimension of the random projection.

While there is, of course, an enormous amount of literature on PCA and related methods, and in addition, there has emerged in more 
recent years a substantial literature on random projection and its integration with various methods for classical problems
(e.g., regression, classification, etc.), to the best of our knowledge there are only two works that, like
ours, explicitly address the use of the tools from these two areas in conjunction with each other.  In the case of the 
first~\cite{papadimitriou2000latent}, a method of random projection followed by subspace projection (via the singular value 
decomposition (SVD)) is proposed for speeding up latent semantic indexing for document analysis.  It is shown~\cite[Thm 5]{papadimitriou2000latent} that, 
with high probability, the result of applying this method to a matrix will yield an approximation of that matrix that is close to what would have been obtained through subspace projection applied to the matrix directly.  A similar result is established in~\cite[Thm 5]{davenport2010signal}, where the goal is to separate a signal of interest from an interfering background signal, under the assumption that the subspace within which either the signal of interest or the interfering signal resides is known.  In both~\cite{papadimitriou2000latent} and~\cite{davenport2010signal}, the proposed methods use a general class of random projections and fixed subspaces.  In contrast, here we restrict our attention specifically to Gaussian random projections but adopt a model-based perspective on the underlying data themselves, specifying that the data derive from a high-dimensional zero-mean multivariate Gaussian distribution with covariance possessed of a compressible set of eigenvalues.  In addition, we study the cases of both known and unknown covariance.  Our results are formulated 
within the context of a hypothesis testing problem and, accordingly, we concentrate on understanding the accuracy with which 
(i) the first two moments of our test statistic is preserved under the null hypothesis, and (ii) the power is preserved under an appropriate alternative hypothesis.
From this perspective, the probabilistic statements in~\cite{papadimitriou2000latent,davenport2010signal} can be interpreted as simpler  precursors
of our results, which nevertheless strongly suggest the feasibility of what we present.
Finally, we note too that the authors in~\cite{davenport2010signal} also propose a method of detection in a hypothesis testing setting, and provide results quantifying the accuracy of power
under random projection, but this is offered separate from their results on subspace projections, and in the context of a model specifying a signal plus
white Gaussian noise.

This paper is organized as follows.  In Section~\ref{sec:bg} we review the standard PCA subspace projection method and
establish appropriate notation for our method of compressed PCA subspace projection.
Our main results are stated in Section~\ref{sec:main.results}, where we
characterize the mean and variance behavior of our statistic $Q^*$ as well as the size and power of
the corresponding statistical test for anomalies based on this statistic.  In Section~\ref{sec:simul.appl}
we present the results of a small simulation study.
 Finally, some brief discussion may be found in Section~\ref{sec:disc}.
The proofs for all theoretical results presented herein may be found in the appendices.

\section{Background}
\label{sec:bg}

Let $X\in \mathbb{R}^{l}$  be a multivariate normal random vector of dimension $l$, with zero mean and
positive definite covariance matrix $\Sigma$. Let $\Sigma=V\Lambda V^T$ be the eigen-decomposition of $\Sigma$.
Denote the prediction of $X$ by the first $k$ principal components of $\Sigma$ as $\hat{X}=(V_kV_k^T)X$.
Jackson and Mudholkar~\cite{jackson.1979}, following an earlier suggestion of Jackson and Morris~\cite{jackson.morris}
in the context of `photographic processing', propose to use the square of the $\ell_2$ norm of the residual from this prediction
as a statistic for testing goodness-of-fit and, more generally, for multivariate quality control.  This is what is referred to now
in the literature as the PCA subspace method.

Denoting this statistic as
\begin{equation}
Q=(X-\hat{X})^T(X-\hat{X}) \enskip ,
\label{eq:Q}
\end{equation}
we know that $Q$ is distributed as a linear combination of independent and identically distributed
chi-square random variables.  In particular,
$$Q\sim\sum_{i=k+1}^l{\sigma_i Z_i^2}\enskip , $$
where $\sigma_i$ are the eigenvalues of $\Sigma$ and the $Z_i$ are independent and identically distributed
standard normal random variables. A normal approximation to this distribution is proposed in~\cite{jackson.1979},
based on a power-transformation and appropriate centering and scaling.
Here, however, we will content ourselves with the simpler approximation of $Q$ by a normal with mean and variance
$$\sum_{i=k+1}^l{\sigma_i} \quad\hbox{and}\quad 2 \sum_{i=k+1}^l \sigma_i^2 \enskip ,$$
respectively.  This approximation is well-justified theoretically (and
additionally has been confirmed in preliminary numerical studies analogous
to those reported later in this paper) by the fact
that $l-k$ typically will be quite large in our context.  In addition, the
resulting simplification will be convenient in facilitating our analysis
and in rendering more transparent the impact of random projection on
our proposed extension of Jackson and Mudholkar's approach.

As stated previously, our extension is motivated by a desire to simultaneously achieve dimension reduction and
ensure data confidentiality.  Accordingly, let $\Phi=(\phi_{ij} )_{l \times p}$,
for $l\gg p$, where the $\phi_{ij}$ are independent and identically distributed
standardized random variables, i.e.,  such that $E(\phi)=0$ and $Var(\phi)=1$.  Throughout this paper we
will assume that the $\phi_{ij}$ have a standard normal distribution. The random matrix $\Phi$ will be used
to induce a random projection
$$\Phi: \mathbb{R}^{l}\to\mathbb{R}^{p},\,\, x\mapsto\frac{1}{\sqrt{p}}\Phi^{T}x \enskip .$$
Note that $\frac{1}{p}\Phi\Phi^T$ tends to the identity matrix $I_{l\times l}$
when $l,p\to\infty$ in an appropriate manner~\cite{bai.yin}.
As a result, we see that
an intuitive advantage of this projection is that the inner product
and the corresponding Euclidean distance are essentially preserved, while reducing
the dimensionality of the space from $l$ to $p$.

Under our intended scenario, rather than observe the original random variable $X$ we instead suppose
that we see only its projection, which we denote as $Y=p^{-1/2}\Phi^T X$.  Consider now the possibility of
applying the PCA subspace method in this new data space.  Conditional on the random matrix $\Phi$,
the random variable $Y$ is distributed as multivariate normal with mean zero and covariance
$\Sigma^{\ast}=(1/p) \Phi^T \Sigma \Phi$.
Denote the eigen-decomposition of this covariance matrix by $\Sigma^{\ast}=U\Lambda^{\ast} U^T$,
let $\hat{Y}=(U_kU_k^T)Y$ represent the prediction of $Y$ by the first $k$ principal components of $\Sigma^{\ast}$,
where $U_k$ is the first $k$ columns of $U$,
and let $\tilde{Y}=Y-\hat{Y}$ be the corresponding residual.  Finally, define the squared $\ell_2$ norm of
this residual as  $$Q^{\ast}=\tilde{Y}^T\tilde{Y}.$$

The primary contribution of our work is to show that, despite not having observed $X$, and therefore being
unable to calculate the statistic $Q$, it is possible, under certain conditions on the covariance $\Sigma$ of $X$ to
apply the PCA subspace method to the projected data $Y$, yielding the statistic $Q^*$, and nevertheless obtain
anomaly detection performance comparable to that which would have been yielded by $Q$, with the discrepancy between the
two made precise.

\section{Main Results}
\label{sec:main.results}

It is unrealistic to expect that the statistics $Q$ and $Q^{\ast}$ would behave comparably under general conditions.
At an intuitive level it is easy to see that what is necessary here
is that the underlying eigen-structure of $\Sigma$ must be sufficiently well-preserved
under random projection.  The relationship between eigen-values and -vectors with and without
random projection is an area that is both classical and the focus of much recent activity.  See
~\cite{bai}, for example, for a recent review.  A popular model in this area is the
{\em spiked covariance model} of Johnstone~\cite{johnstone.2001}, in which it is assumed that
the spectrum of the covariance matrix $\Sigma$ behaves as
$$\sigma_1>\sigma_2\ldots>\sigma_m>\sigma_{m+1}=\ldots=\sigma_l=1 \enskip .$$
This model captures the notion -- often encountered in practice -- of a covariance whose spectrum
exhibits a distinct decay after a relatively few large leading eigenvalues.

All of the results in this section are produced under the assumption of a spiked covariance model.  We present three
sets of results: (i) characterization of the mean and variance of $Q^{\ast}$, in terms of
those of $Q$, in the absence of anomalies;
(ii) a comparison of the power of detecting certain anomalies under $Q^{\ast}$ and $Q$;
and (iii) a quantification of the implications of estimation of $\Sigma^{\ast}$ on our results.

\subsection{Mean and Variance of $Q^{\ast}$ in the Absence of Anomalies}

We begin by studying the behavior of $Q^{\ast}$ when the data are in fact not anomalous, i.e., when
$X$ truly is normal with mean $0$ and covariance $\Sigma$.  This scenario will correspond to the null hypothesis
in the formal detection problem we set up shortly below.  Note that under this scenario,
similar to $Q$, the statistic $Q^{\ast}$ is distributed, conditional on $\Phi$, as a linear combination of
$p-k$ independent and identically distributed chi-square random variables, with mean and variance
given by
$$\sum_{i=k+1}^l{\sigma^{\ast}_i} \quad\hbox{and}\quad 2 \sum_{i=k+1}^l (\sigma^{\ast}_i)^2 \enskip ,$$
respectively, where $(\sigma^{\ast}_1,\ldots,\sigma^{\ast}_p)$ is the spectrum of $\Sigma^{\ast}$.
Our approach to testing will be to first center and scale $Q^{\ast}$, and to then
compare the resulting statistic to a standard normal distribution for testing. 
Therefore, our primary focus in this subsection is on
characterizing the expectation and variance of $Q^{\ast}$

The expectation of $Q^{\ast}$ may be characterized as follows.
\begin{theorem}
\label{thm:Exp.Of.Qstar}
Assume $l,p\to \infty$ such that $\frac{l}{p}=c+o(p^{-1/2})$. If $k>m$
and $\sigma_m>1+\sqrt{c}$, then
\begin{equation}
E_{X|\Phi}(Q^{\ast})=E_X(Q)+O_P(1) \enskip .
\label{eq:Exp.Of.Qstar}
\end{equation}
\end{theorem}
\noindent
Thus $Q^{\ast}$ differs from $Q$ in expectation, conditional on $\Phi$, only by a constant independent
of $l$ and $p$.  Alternatively, if we divide through by $p$ and note that  under the spiked
covariance model
\begin{equation}
\frac{1}{p}\, E_X(Q)=\frac{l-k}{p} \rightarrow c \enskip ,
\label{eq:Exp.Of.Q}
\end{equation}
as $l,p\rightarrow\infty$ , then from (\ref{eq:Exp.Of.Qstar}) we obtain
\begin{equation}
\frac{1}{p}\, E_{X|\Phi}(Q^{\ast})=c+O_P(p^{-1}) \enskip .
\label{eq:Exp.Of.Qstar.p}
\end{equation}
In other words, at the level of expectations, the effect of random projection on  our (rescaled)
test statistic is to introduce a bias that vanishes like $O_P(p^{-1})$.

The variance of $Q^{\ast}$ may be characterized as follows.
\begin{theorem}
\label{thm:Var.Of.Qstar}
Assume $l,p\to \infty$ such that $\frac{l}{p}=c+o(p^{-1/2})$. If $k>m$ and
$\sigma_m>1+\sqrt{c}$, then
\begin{equation}
\frac{\textrm{Var}_{X|\Phi}(Q^{\ast})}{\textrm{Var}_{X}(Q)}= (c+1) +O_P(p^{-1/2}) \enskip .
\label{eq:Var.Of.Qstar}
\end{equation}
\end{theorem}
\noindent
That is, the conditional variance of $Q^*$ differs from the variance of $Q$ by a factor of $(c+1)$, with a
relative bias term of order $O_P(p^{-1/2})$.

Taken together, Theorems~\ref{thm:Exp.Of.Qstar} and~\ref{thm:Var.Of.Qstar}
indicate that application of the PCA subspace method on non-anomalous data
after random projection produces a test statistic
$Q^*$ that is asymptotically unbiased for the statistic $Q$ we would in principle like to use, if
the original data $X$ were available to us, but whose variance is inflated over that of $Q$ by
a factor depending explicitly on the amount of compression inherent in the projection.  In
Section~\ref{sec:simul.appl} we present the results of a small numerical study that show, over a range
of compression values $c$, that the approximations in (\ref{eq:Exp.Of.Qstar.p}) and (\ref{eq:Var.Of.Qstar})
are quite accurate.

\subsection{Comparison of Power for Detecting Anomalies}

We now consider the comparative theoretical performance of the statistics $Q$ and $Q^*$
for detecting anomalies.  From the perspective of the PCA subspace method, an `anomaly' is something that deviates from
the null model that the multivariate normal vector $X$ has mean zero and covariance $\Sigma = V\Lambda V^T$
in such a way that it is visible in the residual subspace, i.e., under projection by $I-V_kV_k^T$.
Hence,we treat the anomaly detection problem in this setting as a hypothesis testing
problem, in which, without loss of generality,
\begin{equation}
H_0: \mu = 0 \quad\hbox{and}\quad H_1: V^T\mu = (\underbrace{0,\ldots,0}_{d>k},\gamma,0,\ldots,0)
\enskip ,
\label{eq:hypotheses}
\end{equation}
for $\mu=E(X)$ and $\gamma > 0$.

Recall that, as discussed in Section~2, it is reasonable in our setting to
approximate the distribution of appropriately standardized versions
of $Q$ and $Q^*$ by the standard normal distribution.
Under our spiked covariance model, and using the results of Theorems~\ref{thm:Exp.Of.Qstar}
and~\ref{thm:Var.Of.Qstar}, this means comparing the statistics
\begin{equation}
\frac{Q-(l -k)}{\sqrt{2(l-k)}} \quad\hbox{and}\quad
\frac{Q^*-(l -k)}{\sqrt{2(l-k)(c+1)}} \enskip ,
\label{eq:std.Q.Qstar}
\end{equation}
respectively, to the upper $1-\alpha$ critical value $z_{1-\alpha}$ of a standard normal distribution.
Accordingly, we define the power functions
\begin{equation}
  \hbox{\sc Power}_Q(\gamma) \, :=\,   \mathbb{P}\left( \frac{Q-(l -k)}{\sqrt{2(l-k)}} > z_{1-\alpha}\right)
\label{eq:Q.power.def}
\end{equation}
and
\begin{equation}
  \hbox{\sc Power}_{Q^*}(\gamma) \, :=\,  \mathbb{P}\left(\frac{Q^*-(l -k)}{\sqrt{2(l-k)(c+1)}} > z_{1-\alpha}\right) \enskip ,
\label{eq:Qstar.power.def}
\end{equation}
for $Q$ and $Q^*$, respectively, where the probabilities $\mathbb{P}$ on the right-hand side of
these expressions refer to the corresponding approximate normal distribution.

Our goal is to understand the relative magnitude of
$\hbox{\sc Power}_{Q^*}$ compared to $\hbox{\sc Power}_Q$, as a function of $\gamma, l, k, c,$ and $\alpha$.
Approximations to the relevant formulas are provided in the following theorem.
\begin{theorem}
Let $Z$ be a standard normal random variable.  Under the same assumptions as Theorems~\ref{thm:Exp.Of.Qstar} and~\ref{thm:Var.Of.Qstar}, and a Gaussian approximation to the standardized test statistics, we have that
\begin{equation*}
\hbox{\sc Power}_Q(\gamma) = \mathbb{P}\left( Z \ge {}_Qz_{1-\alpha}^{crit} \right)
\quad\hbox{and}\quad 
\hbox{\sc Power}_{Q^*}(\gamma) = \mathbb{P}\left( Z \ge {}_{Q^*}z_{1-\alpha}^{crit} \right) \enskip ,
\end{equation*}
where
\begin{equation}
{}_Qz_{1-\alpha}^{crit} = \frac{z_{1-\alpha}\sqrt{2(l-k)} - \gamma^2}
                                          {\sqrt{2(l-k) + 4\gamma^2}}
\label{eq:Q.power.eqn}
\end{equation}
while
\begin{equation}
{}_{Q^*}z_{1-\alpha}^{crit} = \frac{z_{1-\alpha}\sqrt{2(l-k)} - \left[ \gamma^2/\sqrt{c+1} + O_P(1)\right]}
                                                 {\sqrt{2(l-k) + 4\gamma^2 + O_P(p^{1/2})}} \enskip .
\label{eq:Qstar.power.eqn}
\end{equation}
\label{thm:power}
\end{theorem}

Ignoring error terms, we see that the critical values (\ref{eq:Q.power.eqn}) and (\ref{eq:Qstar.power.eqn}) for both power formulas
have as their argument quantities of the form $c_1 z_{1-\alpha} - c_2$.  However, while $c_1(Q^*)\approx c_1(Q)$, we have that
$c_2(Q^*) \approx c_2(Q)/(c+1)^{1/2}$.  Hence, all else being held equal, as the compression ratio $c$ increases,
the critical value at which power is evaluated shifts increasingly to the right for $Q^*$, and power decreases
accordingly.  The extent to which this effect will be apparent is modulated by the magnitude $\gamma$ of the
anomaly to be detected and the significance level $\alpha$ at which the test is defined, and furthermore by
the size $l$ of the original data space.  Finally, while these observations can be expected to be most accurate for
large $l$ and large $\gamma$, in the case that either or both are more comparable in size to the $O_P(p^{1/2})$ and $O_P(1)$ error 
terms in (\ref{eq:Qstar.power.eqn}), respectively, the latter will play an increasing role and hence affect the accuracy of 
the stated results.

\begin{figure}[bht]
\vspace{6pc}
\caption[]{$\hbox{\sc Power}_{Q^*}$ as a function of compression ratio $c$.}
\begin{center}
\includegraphics[width=0.75\textwidth]{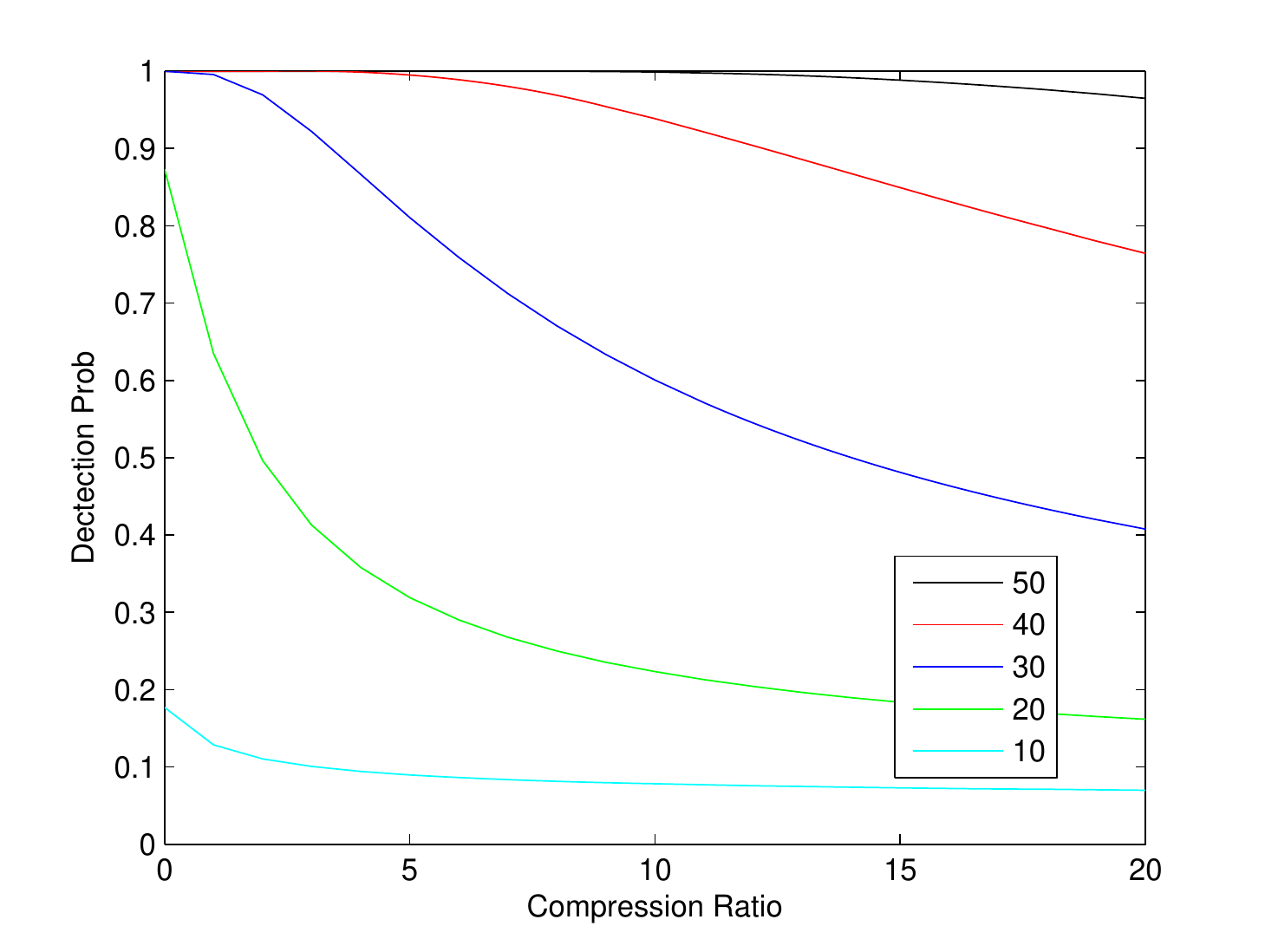}
\end{center}
\label{fig:power}
\end{figure}

An illustration may be found in Figure~\ref{fig:power}.  There we show the power $\hbox{\sc Power}_{Q^*}$ as a function of the compression ratio $c$, for $\gamma=10,20,30,40$, and $50$.  Here the dimension before projection is $l=10,000$ and the dimension after projection $p=l/c$ ranges from $10,000$ to $500$.  A value of $k=30$ was used for the dimension of the principle component analysis, and a choice of $\alpha=0.05$ was made for the size of the underlying test for anomaly.  Note that at $c=0$, on the far left-hand side of the plot, the value $\hbox{\sc Power}_{Q^*}$ simply reduces to $\hbox{\sc Power}_Q$.  So the five curves show the loss of power resulting from compression, as a function of compression level $c$, for various choices of strength $\gamma$ of the anomaly.

Additional numerical results of a related nature are presented in Section~\ref{sec:simul.appl}. 

\subsection{Unknown Covariance}

The test statistics $Q$ and $Q^*$ are defined in terms of the covariance matrices $\Sigma$ and $\Sigma^*$,
respectively.  However, in practice, it is unlikely that these matrices are known.  Rather, it is more likely
that estimates of their values be used in calculating the test statistics, resulting, say, in statistics
$\hat Q$ and $\hat Q^*$.  In the context of industrial systems control, for example, it is not unreasonable
to expect that there be substantial previous data that may be used for this purpose.
As our concern in this paper is on the use of the subspace projection method after random projection, i.e.,
in the use of $Q^*$, the relevant question to ask here is what are the implications of using an estimate
$\hat\Sigma^*$ for $\Sigma^*$.  

We study the natural case where the estimate $\hat\Sigma^*$
is simply the sample covariance $\frac{1}{n}(\mathbf{Y}-\bar{Y})(\mathbf{Y}-\bar{Y})^T$, for
$\mathbf{Y} = [Y_1,\ldots,Y_n]$ the $p\times n$ matrix formed from
$n$ independent and identically distributed copies of the random variable $Y$ and $\bar{Y}$ their
vector mean.  Let $\hat U \hat\Lambda^* \hat U^T$ be the eigen-decomposition of $\hat\Sigma^*$ and, accordingly,
define $\hat{Q}^{\ast}=Y^T(I-\hat{U}_k\hat{U}_k^T)Y$ in analogy to $Q^{\ast}=Y^T(I-U_k U_k^T)Y$.
We then have the following result.
\begin{theorem}
Assume $n\ge p$.  Then, under the same conditions as Theorem~\ref{thm:Exp.Of.Qstar},
\begin{equation}
E_{X|\Phi}(\hat Q^{\ast})=E_X(Q)+O_P(1) 
\label{eq:Exp.Of.Qhatstar}
\end{equation}
and
\begin{equation}
\frac{\textrm{Var}_{X|\Phi}(\hat Q^{\ast})}{\textrm{Var}_{X}(Q)}= (c+1) +O_P(p^{-1/2}) \enskip .
\label{eq:Var.Of.Qhatstar}
\end{equation}
Furthermore, under the conditions of Theorem~\ref{thm:power}, the power function
\begin{equation}
  \hbox{\sc Power}_{\hat Q^*}(\gamma) \, :=\,  \mathbb{P}\left(\frac{\hat Q^*-(l -k)}{\sqrt{2(l-k)(c+1)}} > z_{1-\alpha}\right) 
\label{eq:Qstarhat.power.def}
\end{equation}
can be expressed as $\mathbb{P}\left( Z \ge {}_{\hat Q^*}z_{1-\alpha}^{crit} \right)$,
where
\begin{equation}
{}_{\hat Q^*}z_{1-\alpha}^{crit} = \frac{z_{1-\alpha}\sqrt{2(l-k)} - \left[ \gamma^2/\sqrt{c+1} + O_P(1)\right]}
                                                 {\sqrt{2(l-k) + 4\gamma^2 + O_P(p^{1/2})}} \enskip .
\label{eq:Qstarhat.power.eqn}
\end{equation}
\label{thm:Qhatstar}
\end{theorem}

Simply put, the results of the theorem tell us that the accuracy with which compressed PCA subspace projection approximates standard PCA subspace projection
in the original data space, when using the estimated covariance $\hat\Sigma^*$ rather than the unknown covariance $\Sigma^*$, is unchanged, as long
as the sample size $n$ used in computing $\hat\Sigma^*$ is at least as large as the dimension $p$ after random projection.  Hence, there is an interesting trade off
between $n$ and $p$, in that the smaller the sample size $n$ that is likely to be available, the smaller the dimension $p$ that must be used in defining our random projection, if the ideal accuracy is to be obtained (i.e., that using the true $\Sigma^*$).  However, decreasing $p$ will degrade the quality of the accuracy in this ideal case, as it increases the compression parameter $c$.

\section{Simulation}
\label{sec:simul.appl}

We present two sets of numerical simulation results in this section, one corresponding to Theorems~\ref{thm:Exp.Of.Qstar}
and~\ref{thm:Var.Of.Qstar}, and the other, to Theorem~\ref{thm:power}.

In our first set of experiments, we simulated from the spiked covariance model, drawing both random variables $X$
and their projections $Y$ over many trials, and computed $Q$ and $Q^*$ for each trial, thus allowing us to compare
their respective means and variances.  In more detail, we let the dimension
of the original random variable $X$ be $l=10,000$, and assumed that to be distributed as normal with mean zero
and (without loss of generality) covariance equal to the spiked spectrum
$$\sigma_1=50,\sigma_2=40,\sigma_3=30,\sigma_4=20,\sigma_5=10,\sigma_6=\ldots=\sigma_l=1 \, ,$$
with $m=5$.  The corresponding random projections $Y$ of $X$ were computed using random matrices $\Phi$
generated as described in the text, with  compression ratios $c=l/p$ equal to $20,50,$ and $100$ (i.e., $p=500,200,$ and $100$).  We used a total of  $2000$ trials for each realization of $\Phi$, and $30$ realizations of $\Phi$ for each
choice of $c$ ($p$).

The results of this experiment are summarized in Table~\ref{tab:Exp1}.  Recall that Theorems~\ref{thm:Exp.Of.Qstar} and~\ref{thm:Var.Of.Qstar} say that the rescaled mean $E(Q^\ast)/p$ and the ratio of variances $Var({Q^\ast})/Var(Q)$
should be approximately equal to $c$ and $c+1$, respectively.  It is clear from these results that, for low levels of compression (i.e., $c=20$) the approximations
in our theorems are quite accurate and that they vary little from one projection to another.  For moderate levels
of compression (i.e., $c=50$) they are similarly accurate, although more variable.  For high levels of compression (i.e., $c=100$), we begin to see some non-trivial bias entering, with some accompanying increase in variability as well.
\begin{table}[hbt]
\begin{center}
\begin{tabular}{|c|c|c|c|}
\hline
c  & p  & $ E({Q^*})/p$ & $Var({Q^*})/Var(Q)$ \\
\hline
20 & 500  &19.681(0.033)  &20.903(0.571) \\
50 & 200  &48.277(0.104)  &50.085(1.564)  \\
100& 100 &93.520(0.346)  &96.200(3.871)  \\
\hline
\end{tabular}
\end{center}
\caption{Simulation results assessing the accuracy of Theorems~\ref{thm:Exp.Of.Qstar} and~\ref{thm:Var.Of.Qstar}.
\label{tab:Exp1}}
\end{table}

In our second set of experiments, we again simulated from a spiked covariance 
model, but now with non-trivial mean.  The spiked spectrum was chosen to be
the same as above, but with $l=5000$, for computational considerations.
The mean was defined as in (\ref{eq:hypotheses}), with $\gamma = 20, 30,$
or $40$.  A range of compressions ratios $c=1,2,\ldots, 20$ were used.
We ran a total of $1000$ trials for each realization of $\Phi$, 
and $30$ realizations of $\Phi$ for each combination of $c$ and $\gamma$.
The statistics $Q$ and $Q^*$ were computed as in the statement of 
Theorem~\ref{thm:power} and compared to the critical value 
$z_{0.95}=1.645$, corresponding to a one-sided test of size $\alpha=0.05$.

\begin{figure}[bht] 
\caption[]{Simulation results assessing the accuracy 
            of Theorem~\ref{thm:power}.}
\begin{center}
\includegraphics[width=0.75\textwidth]{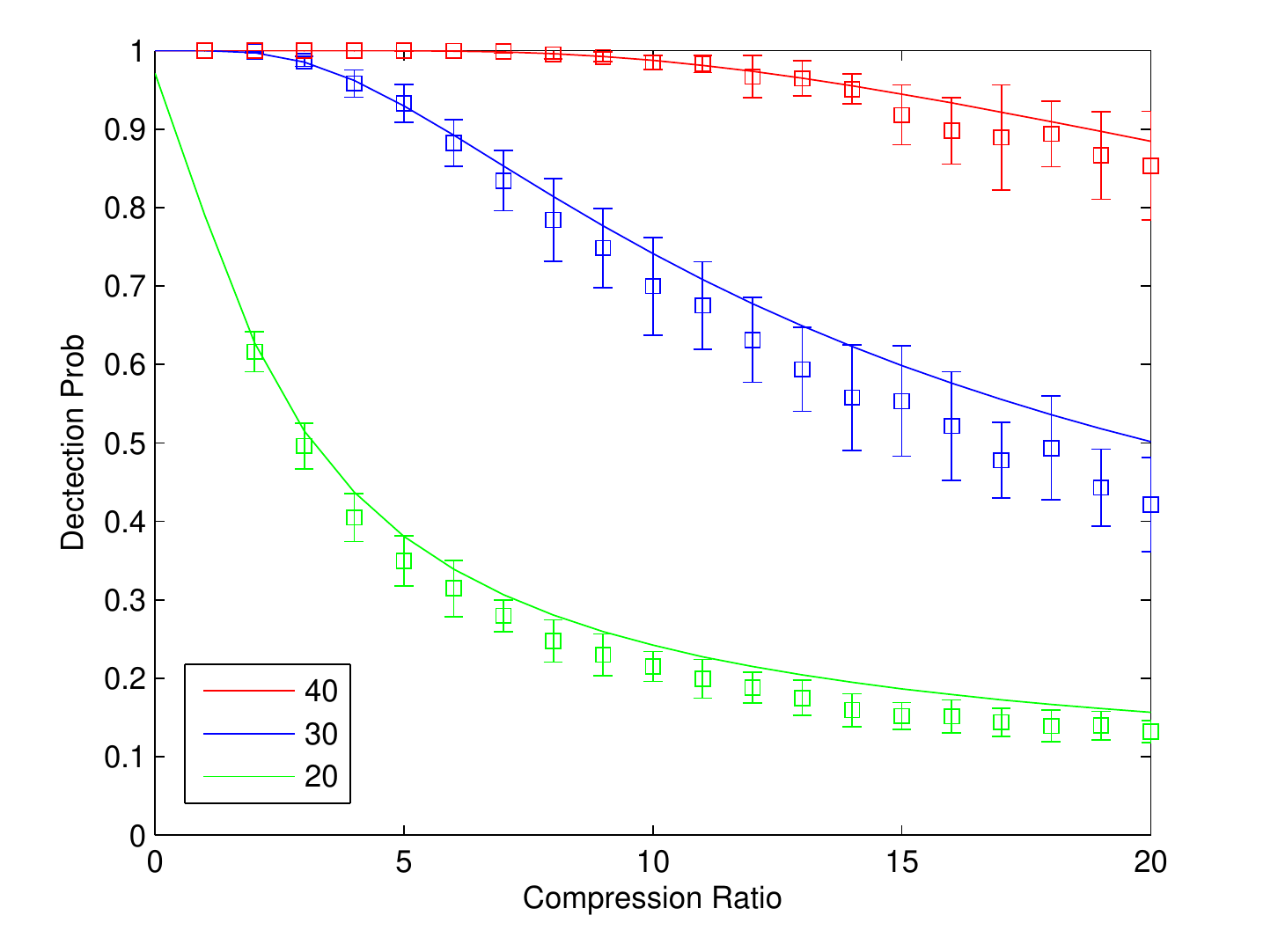}
\end{center}
\label{fig:simul_power}
\end{figure}

The results are shown in Figure~\ref{fig:simul_power}.  Error bars reflect
variation over the different realizations of $\Phi$ and correspond to one
standard deviation.  The curves shown correspond to the power approximation
$\hbox{\sc Power}_{Q^*}$ given in Theorem~\ref{thm:power}, and are the same as
the middle three curves in Figure~\ref{fig:power}.  We see that for the 
strongest anomaly level ($\gamma=40$) the theoretical approximation matches the 
empirical results quite closely for all but the highest levels of compression.
Similarly, for the weakest anomaly level ($\gamma=20$), the match is also 
quite good, although there appears to be a small but persistent positive bias
in the approximation across all compression levels.  In both cases, the 
variation across choice of $\Phi$ is quite low.  The largest bias in the
approximation is seen at the moderate anomaly level ($\gamma=30$), at
moderate to high levels of compression, although the bias appears to be
on par with the anomaly levels at lower compression levels.  The largest
variation across realizations of $\Phi$ is seen for the moderate anomaly level.

\section{Discussion}
\label{sec:disc}
Motivated by dual considerations of dimension reduction and 
data confidentiality, as well as the wide-ranging and successful 
implementation of PCA subspace projection, we have introduced a 
method of {\em compressed} PCA subspace projection and characterized
key theoretical quantities relating to its use as a tool in anomaly 
detection.  An implementation of this proposed methodology and its
application to detecting IP-level volume anomalies in computer network traffic
suggests a high relevance to practical problems~\cite{hawk.thesis}.
Specifically, numerical results generated using archived Internet traffic data 
suggest that, under reasonable levels of compression $c$,  it is possible to detect volume-based anomalies (i.e., in units of
bytes, packets, or flows) using compressed PCA subspace detection at almost $70\%$ the power of the uncompressed method.

The results of Theorem~\ref{thm:Qhatstar} are important in establishing the practical feasibility of our proposed method, wherein 
the covariance $\Sigma^*$ must be estimated from data, when it is possible to obtain samples of size $n$ of a similar order of magnitude 
as the reduced dimension $p$ of our random projection.  It would be of interest to establish results of a related nature for the case where $n\ll p$.
In that case, it cannot be expected that the classical moment-based estimator $\hat\Sigma^*$ that we have used here will perform acceptably.
Instead, an estimator exploiting the structure of $\Sigma^*$ presumably is needed.  However, as most methods in the recent literature on 
estimation of large, structured covariance matrices assume sparseness of some sort
(e.g., \cite{bickel2008regularized,karoui2008operator,levina2008sparse}), they are unlikely to be applicable here, since $\Sigma^*$ is
roughly of the form $c I_{p\times p} + W$, where $W$  is of rank $m$ with entries of magnitude $o_P(p^{-1})$.  
Similarly, neither will methods of sparse PCA be appropriate (e.g, \cite{zht,johnstone.lu,shen2008sparse,journee2010generalized}).
Rather, variations on more recently proposed methods aimed directly at capturing low-rank covariance structure hold promise (e.g., \cite{negahban2011estimation,luo2011high}).  
Alternatively, the use of so-called very sparse random projections
(e.g., \cite{li2006very}), in place of our Gaussian random projections, would yield sparse covariance matrices $\Sigma^*$, and hence in principle facilitate
the use of sparse inference methods in producing an estimate $\hat\Sigma^*$.  But this step would likely come at the cost of making the already fairly detailed
technical arguments behind our results more involved still, as we have exploited the Gaussianity of the random projection in certain key places to simplify
calculations.  We note that ultimately, for such approaches to produce results of accuracy similar to that here in Theorem~4, it is necessary that they produce
approximations to the PCA subspace of $\Sigma^*$ with order $O_P(n^{-1/2})$ accuracy.

Finally, we acknowledge that the paradigm explored here, based on Gaussian random
projections, is only a caricature of what might be implemented in reality,
particularly in contexts like computer network traffic monitoring.  There,
issues of data management, speed, etc. would become important and can be
expected to have non-trivial implications on the design of the type of
random projections actually used.  Nevertheless, we submit that the
results presented in this paper strongly suggest the potential success of an 
appropriately modified system of this nature.

\section{Appendix}

\subsection{Proof of Theorem 1}

Suppose random vector $X\in\mathbb{R}^l$ has a multivariate
Gaussian distribution $N(0,\Sigma_{l\times l})$ and $Y\sim
N(0,\Sigma^{\ast}_{p\times p})$, for $\Sigma^{\ast}=\frac{1}{p}\Phi'\Sigma\Phi$.
Denote the eigenvalues of $\Sigma$ and $\Sigma^{\ast}$ as
$(\sigma_1,\ldots,\sigma_l)$ and
$(\sigma^\ast_1,\ldots,\sigma^\ast_p)$, respectively.

Jackson and Mudholkar~\cite{jackson.1979} show that
$Q=(X-\hat{X})'(X-\hat{X})$ will be distributed as
$\sum\limits_{i=k+1}^{l}\sigma_iZ_i^2$, where the  $Z_i$ are
independent and identically distributed (i.i.d.) standard normal random
variables. Consequently, we have
$E_X(Q)=\sum\limits_{i=k+1}^{l}\sigma_i$ and, similarly,
$E_{X|\Phi}(Q^{\ast})=\sum\limits_{i=k+1}^{p}\sigma^\ast_i$. So
comparison of $E_X(Q)$ and $E_{X|\Phi}(Q^{\ast})$ reduces to a comparison
of partial sums of the eigenvalues of $\Sigma$ and
$\Sigma^{\ast}$.

Since
$$E_X(Q)=\sum_{i=k+1}^{l}\sigma_i=\sum_{i=1}^{l}\sigma_i-\sum_{i=1}^{k}\sigma_i=tr(Q)-\sum_{i=1}^{k}\sigma_i \enskip ,$$
in the following proof we will analyze $tr(Q)$ and
$\sum\limits_{i=1}^{k}\sigma_i$ separately.

\subsubsection{}
Because orthogonal rotation has no influence on Gaussian random
projection and the matrix spectrum, to simplify the computation, we
assume without loss of generality that $\Sigma=diag(\sigma_1,\sigma_2,\ldots,\sigma_l)$. So the
diagonal elements of $\Sigma^{\ast}=\frac{1}{p}\Phi^T\Sigma\Phi$
are
$$\sigma_{jj}^{\ast}=\frac{1}{p}\sum_{i=1}^{l}{\phi_{ij}^2\sigma_i}$$

We have
$$tr(\Sigma^{\ast})=\frac{1}{p}\sum^p_{j=1}
(\sum^l_{i=1}\phi_{ij}^2\sigma_i)=\sum^l_{i=1}
\sigma_i(\frac{1}{p}\sum^p_{j=1}\phi_{ij}^2) \quad\hbox{and}\quad
tr(\Sigma)=\sum^l_{i=1}\sigma_i$$
and therefore
$$tr(\Sigma^{\ast})-tr(\Sigma)=\sum^l_{i=1}\sigma_i(\frac{1}{p}\sum^p_{j=1}\phi_{ij}^2-1).$$

Under the \textit{spiked covariance model} assumed in this paper, $\sigma_1>\sigma_2>\ldots>\sigma_m>\sigma_{m+1}=\sigma_{m+2}=\ldots=\sigma_l=1$
for fixed $m$. Then,
$$tr(\Sigma^{\ast})-tr(\Sigma)=\sum^m_{i=1}\sigma_i(\frac{1}{p}\sum^p_{j=1}\phi_{ij}^2-1)+\sum^l_{i=m+1}(\frac{1}{p}\sum^p_{j=1}\phi_{ij}^2-1).$$
When $l,p\to \infty,\frac{l}{p}\to c$ and the first term will go to
zero like $O_P(p^{-1/2})$.  The second term can be written
as:
$$(l-m)\frac{1}{(l-m)p}\sum^l_{i=m+1}\sum^p_{j=1}(\phi_{ij}^2-1)$$

More precisely, here we have a series $\{l_n\},\{p_n\}$
satisfying $l_n\to\infty,\, p_n\to\infty,\, \frac{l_n}{p_n}\to c>0$ when
$n\to\infty$. It is easy to show that $D_n=(l_n-m)p_n\to \infty$
when $n\to \infty$.  Since the $\phi_{ij}$ are i.i.d., we can
re-express the series
$\frac{1}{(l_n-m)p_n}\sum\limits^l_{i=m+1}\sum\limits^{p_n}_{j=1}(\phi_{ij}^2-1)$
as
$$\frac{1}{D_n}\sum^{D_n}_{i'=1}(\phi_{i'}^2-1)$$

Recalling that the $\phi$ are standard normal random variables, we
know that $E(\phi^2-1)=0$ and $Var(\phi^2-1)=2$. By the central limit theorem, the series
$\{\sqrt{N}\frac{1}{N}\sum\limits^N_{i''=1}(\phi_{i''}^2-1)\}^\infty_{N=1}$
will converge to a zero mean normal in distribution. Hence
$\frac{1}{N}\sum\limits^N_{i''=1}(\phi_{i''}^2-1)$ is of  order
$O_P(N^{-1/2})$ when $N\to\infty$. As an infinite
subsequence,
$$\frac{1}{D_n}\sum^{D_n}_{i'=1}(\phi_{i'}^2-1)$$
also has the same behavior, which leads to
$$\frac{1}{D_n}\sum^{D_n}_{i'=1}(\phi_{i'}^2-1)=O_P(D_n^{-1/2})=O_P([(l_n-m)p_n]^{-1/2}) \enskip ,$$
by which we conclude that
$$(l-m)\frac{1}{(l-m)p}\sum^l_{i=m+1}\sum^p_{j=1}(\phi_{ij}^2-1)=(l-m)O_P([(l-m)p]^{-1/2})=O_P(1).$$

As a result of the above arguments,
$$tr(\Sigma^{\ast})-tr(\Sigma)=O_P(p^{-1/2})+O_P(1)=O_P(1).$$

\subsubsection{}

Next we examine the behavior of the first $k$ eigenvalues of $\Sigma$ and
$\Sigma^{\ast}$, i.e.,  $\{\sigma_1\ldots\sigma_k\}$ and
$\{\sigma_1^{\ast}\ldots\sigma_k^{\ast}\}$.

Recalling the definition of $Y$ as $Y=\frac{1}{\sqrt{p}}\Phi'X\sim N(0,\frac{1}{p}\Phi^T\Sigma\Phi)$,
we define the ${l\times p}$ matrix $Z=\Sigma^{1/2}\Phi$.  All of the
columns of $Z$ are i.i.d random vectors from $N(0,\Sigma)$, and
$\frac{1}{p}\Phi^T\Sigma\Phi$, the covariance of $Y$, can be expressed as $\frac{1}{p}Z'Z$.  Let
$S=\frac{1}{p}ZZ'$, which contains the same non-zero eigenvalues as
$\Sigma^{\ast}=\frac{1}{p}Z'Z$. Through this transformation of $Y$ to $Z$
and interpretation of $S$ as the sample covariance corresponding to $\Sigma$, we are able
to utilize established results from random matrix theory.

Denote the spectrum of $S$ as
$(s_1,\ldots,s_p)$.  Under the spiked covariance model, Baik~\cite{bai} and Paul~\cite{debashis}
independently derived the limiting behavior of the elements of this spectrum.  Under our
normal case $Z_i\sim N(0,\Sigma)$, Paul~\cite{debashis} proved the
asymptotical normality of $s_v$.
\begin{theorem}
Assume $l,p\to \infty$ such that $\frac{l}{p}=c+o(p^{-1/2})$. If
$\sigma_v>1+\sqrt{c}$, then
$$\sqrt{p}(\sigma_v^\ast-\sigma_v(1+\frac{c}{\sigma_v-1}))\Rightarrow N(0,2\sigma_v^2(1-\frac{c}{(\sigma_v-1)^2}))
\enskip . $$
\end{theorem}

For significantly large leading eigenvalues $\sigma_v\gg1$, $s_v$
is asymptotically $N(\sigma_v,\frac{2}{p}\sigma_v^2)$. And for all
of the lead eigenvalues which are above the threshold
$1+\sqrt{c}$, we have
$\sigma_v^\ast-\sigma_v=O_P(p^{-1/2})$.  Recalling the condition
$k\ge m$ in the statement of the theorem, without loss of generality we
take $k=m$ (as we will do, when convenient, throughout the rest of the proofs in these appendices).
Using Paul's result, we have
$$\sum_{i=1}^{k}{\sigma^{\ast}_i}=\sum_{i=1}^{k}{\sigma_i}+O_P(p^{-1/2})$$

Combining these results with those of the previous subsection, we have
$$E_{X|\Phi}(Q^{\ast})-E_X(Q)=(tr(\Sigma^{\ast})-tr(\Sigma))+(\sum_{i=1}^{k}{\sigma^{\ast}_i}-\sum_{i=1}^{k}{\sigma_i})=O_p(1) \enskip .$$


\subsection{Proof of Theorem 2}
For notational convenience, denote
$\Sigma^{\ast}$ as $(A_{ij})_{p\times p}$, so that
$\|\Sigma^{\ast}\|_F^2=\sum{(A_{ij}^2)}$.  Writing
$Q=\sum\limits_{i=k+1}^{l}\sigma_i Z_i^2$, and similarly for $Q^*$, we have
$$\textrm{Var}_X(Q)=2(\sum_{i=k+1}^l\sigma_i^2) \quad\hbox{and}\quad
\textrm{Var}_{X|\Phi}(Q^{\ast})=2(\sum\limits_{i=k+1}^p{\sigma^{\ast}_i}^2)=2(\|\Sigma^{\ast}\|_F^2-\sum\limits_{i=1}^k{\sigma^{\ast}_i}^2)   \enskip .$$

Since $\Sigma^{\ast}=\frac{1}{p}\Phi^T\Sigma\Phi$, we have
$$A_{ij}=\frac{1}{p}\sum_{h=1}^l\phi_{ih}\phi_{jh}\sigma_h \enskip .$$
Accodingly, if $i=j$,
$$A_{ii}=\frac{1}{p}\sum_{h=1}^{l}\phi_{ih}^2\sigma_h$$
and
$$A_{ii}^2=\frac{1}{p^2}\left[\sum_{h=1}^{l}\phi_{ih}^4\sigma_h^2+\sum_{h\neq h'}\phi_{ih}^2\phi_{ih'}^2\sigma_h\sigma_{h'}
  \right]  \enskip ,$$
while if $i\neq j$,
$$A_{ij}^2=\frac{1}{p^2}\left[\sum_{h=1}^l\phi_{ih}^2\phi_{jh}^2\sigma_h^2+\sum_{h\neq
h'}\phi_{ih}\phi_{jh}\phi_{ih'}\phi_{jh'}\sigma_h\sigma_{h'} \right] \enskip .$$

Changing the order of summation, we therefore have
$$\|\Sigma^{\ast}\|_F^2=\frac{1}{p^2}\left[\sum_{h=1}^l\sigma_h^2 \left(\sum_{i=1}^p\phi_{ih}^4+\sum_{i\neq
j}\phi_{ih}^2\phi_{jh}^2 \right)+\sum_{h\neq
h'}\sigma_h\sigma_{h'} \left(\sum_{i=1}^p\phi_{ih}^2\phi_{ih'}^2+\sum_{i\neq
j}\phi_{ih}\phi_{jh}\phi_{ih'}\phi_{jh'}\right) \right]$$
which implies,
\begin{equation}
\|\Sigma^{\ast}\|_F^2=\sum_{h=1}^l\sigma_h^2 \left(\frac{1}{p}\sum_{i=1}^p\phi_{ih}^2\right)^2+\sum_{h\neq
h'}\sigma_h\sigma_{h'}\left(\frac{1}{p}\sum_{i=1}^p\phi_{ih}\phi_{ih'}\right)^2 \enskip .\label{B1}
\end{equation}

Now under the spiked covariance model, with $k=m$, we have
$$\textrm{Var}(Q)=2 \left(\sum\limits_{h=k+1}^l\sigma_h^2 \right)=2\left(l-m\right)
  \enskip .$$
As a result,  we have
$$\frac{\textrm{Var}_{X|\Phi}(Q^{\ast})}{\textrm{Var}(Q)}=\frac{\|\Sigma^{\ast}\|_F^2-\sum\limits_{h=1}^{k}{\sigma^\ast_h}^2}{\sum_{i=k+1}^l\sigma_i^2} = \frac{1}{l-m}\left(\|\Sigma^{\ast}\|_F^2-\sum\limits_{h=1}^{k}{\sigma^\ast_h}^2\right) \enskip .$$
Substituting the expression in equation \ref{B1} yields
\begin{eqnarray}
\frac{\textrm{Var}_{X|\Phi}(Q^{\ast})}{\textrm{Var}(Q)} & = & \frac{1}{l-m} \left[\sum_{h=1}^l\sigma_h^2 \left(\frac{1}{p}\sum_{i=1}^p\phi_{ih}^2\right)^2
-\sum_{h=1}^{k}{\sigma^\ast_h}^2 \right] \nonumber \\
 &  & +\frac{1}{l-m}{\sum_{h\neq h'}\sigma_h\sigma_{h'} \left(\frac{1}{p}\sum_{i=1}^p\phi_{ih}\phi_{ih'}\right)^2}.\label{B2}
\end{eqnarray}

The control of equation \ref{B2} is not immediate. Let us
denote the two terms in the RHS of \ref{B2} as $A$ and $B$. Results in
the next two subsections show that $A$ behaves like
$1+O_P(p^{-1/2})$, and $B$, like $c+O_P(p^{-1/2})$.
Consequently, Theorem~2 holds.

\subsubsection{}

We show in this subsection that
$$A=\frac{1}{l-m}\left[\sum_{h=1}^l\sigma_h^2 \left(\frac{1}{p}\sum_{i=1}^p\phi_{ih}^2\right)^2
-\sum_{h=1}^{k}{\sigma^\ast_h}^2 \right]=1+O_P(p^{-1/2}) \enskip .$$

First note that, by an appeal to the central limit theorem,
$\frac{1}{p}\sum\phi_{ih}^2=1+O_P(p^{-1/2})$.
So A can be expressed as
$$\frac{1}{l-m}\left\{\left(\sum_{h=1}^k\sigma_h^2-\sum_{h=1}^k{\sigma^\ast_h}^2\right)
+\sum_{h=1}^k \left[\sigma_h^2O_P(p^{-\frac{1}{2}}) \right]
 +\sum_{h=k+1}^l\sigma_h^2 \left[1+O_P(p^{-\frac{1}{2}})\right] \right\}  \enskip .$$
Using the result by Paul cited in Section~A.2, in the form of Theorem 6,
the first term is found to behave like $O_P(p^{-1})$.  In addition, it easy to see that the
second term behaves like $O_P(p^{-3/2})$.  Finally, since under the spiked
covariance model $\sigma_{m+1}=\ldots=\sigma_l=1$, taking $k=m$ we have that
$$\sum\limits_{h=k+1}^l\sigma_h^2\left[1+O_P(p^{-\frac{1}{2}})\right]
  =(l-m)\left[1+O_P(p^{-1/2})\right] \enskip .$$
As a result, the third term in the expansion of $A$ is equal to
$1+O_P(p^{-1/2})$.

Combining terms, we find that $A=1+O_P(p^{-1/2})$.

\subsubsection{}
Term B in \ref{B2} can be written as
\begin{equation}
B=
\frac{2}{p^2(l-m)}{\sum_{1\leq h'< h\leq l}
\sigma_h\sigma_{h'}(\sum_{i=1}^p\phi_{ih}\phi_{ih'})^2}\label{B3} \enskip .
\end{equation}
Recalling that under the spiked covariance model
$\sigma_1>\sigma_2\ldots>\sigma_m>\sigma_{m+1}=\ldots=\sigma_l=1$,
in the following we will analyze the asymptotic behavior of the term B in two stages,
by first handling the case $\sigma_1=\sigma_2=\ldots=\sigma_l=1$ in detail,
and second, arguing that the result does not change under the original conditions.

If $\sigma_1=\sigma_2=\ldots=\sigma_l=1$, which is simply a white noise model, term B becomes,
\begin{equation}
\frac{2}{p^2(l-m)}{\sum_{1\leq h'< h\leq l}
\left(\sum_{i=1}^p\phi_{ih}\phi_{ih'}\right)^2} \enskip ,
\end{equation}
which may be usefully re-expressed as
\begin{equation}
\frac{2}{p^2(l-m)}{\sum_{1\leq h'< h\leq l}}
\left(\sum\limits_{i=1}^p\phi_{ih}^2\phi_{ih'}^2+2\sum\limits_{i>j}\phi_{ih}\phi_{ih'}\phi_{jh}\phi_{jh'}\right)
\enskip ,
\end{equation}
and, upon exchanging the order of summation, as
\begin{equation}
\frac{2}{p^2(l-m)}\sum_{i=1}^p \sum_{1\leq h'< h\leq l}\phi_{ih}^2\phi_{ih'}^2 \,
+\, \frac{4}{p^2(l-m)}\sum_{i>j}\sum_{1\leq h'< h\leq l}
\phi_{ih}\phi_{ih'}\phi_{jh}\phi_{jh'} \enskip .\label{B6}
\end{equation}

Write equation \ref{B6} as $B=B_1+B_2$.  In the material that immediately follows,
we will argue that, under the conditions of the theorem and the white noise model,
$B_1=c+O_P(p^{-1/2})$ and $B_2=O_P(p^{-1})$.

To prove the first of these two expressions, we begin by writing
$$T_i = \sum\limits_{h>h'}^l\phi_{ih}^2\phi_{ih'}^2 \quad\hbox{and}\quad
B_1 = \frac{2}{p(l-m)}\left(\frac{1}{p}\sum\limits_{i=1}^pT_i\right) \enskip .$$
Note that the $T_i$ are i.i.d. random variables.  We will use a central limit theorem
argument to control $B_1$.

A straightforward calculation shows that $E(T_i)= l(l-1)/2$.  To characterize the second moment, we write
$$T_i^2= \left(\sum\limits_{1\leq h'<h\leq
l}\phi_{ih}^2\phi_{ih'}^2\right)^2=\sum\limits_{h'<h;H'<H}\phi_{ih}^2\phi_{ih'}^2\phi_{iH}^2\phi_{iH'}^2$$
and consider each of three possible types of terms
$\phi_{ih}^2\phi_{ih'}^2\phi_{iH}^2\phi_{iH'}^2$ \, .
\begin{enumerate}
  \item If $H=h,H'=h'$, then $\phi_{ih}^2\phi_{ih'}^2\phi_{iH}^2\phi_{iH'}^2=\phi_{ih}^4\phi_{ih'}^4$,
with expectation $9$.  Since there are $l(l-1)/2$ such
choices of $(h,h')$, the contribution of  terms from this case to $E(T_i^2)$ is
$9[l(l-1)]/2$.

  \item If only two of $(h,h',H,H')$ are equal,
$\phi_{ih}^2\phi_{ih'}^2\phi_{iH}^2\phi_{iH'}^2$ will take the form
$\phi_{ia}^2\phi_{ib}^2\phi_{ic}^4$ with expectation 3. For each
triple $(a,b,c)$ there are six possible cases:
$h=H>h>H'$,$h=H>H'>h$,$h>h'=H>H'$,$H>H'=h>h'$,$H>h>H'=h'$,$h>H>H'=h'$.
So there are $l(l-1)(l-2)$ such terms in this case, yielding a contribution of $3l(l-1)(l-2)$
to $E(T_i^2)$.

\item If $(h,h',H,H')$ are all different, the expectation of
$\phi_{ih}^2\phi_{ih'}^2\phi_{iH}^2\phi_{iH'}^2$ is just 1. Since
there are $\frac{l^2(l-1)^2}{4}$ terms in total, the number of such terms in this case
and hence the contribution of this case to $E(T_i^2)$ is
$\frac{l^2(l-1)^2}{4}-\frac{l(l-1)}{2}-l(l-1)(l-2)$.
\end{enumerate}
Combining these various calculations we find that
$$E(T_i^2)=\frac{l^2(l-1)^2}{4}+8\,\frac{l(l-1)}{2}+2l(l-1)(l-2)$$
and hence
$$\textrm{Var}(T_i)=E(T_i^2)-E(T_i)^2= 2l^2 (l-1)\enskip .$$

By the central limit theorem we know that $\sqrt{p}\left(\bar{T} - E[T]\right)/\sqrt{Var(T)} = O_P(1)$.
Exploiting that $B_1 = [2/p(l-m)] \bar{T}$ and recalling that $l/p = c + o(p^{-1/2})$ by assumption,
simple calculations yield that $B_1 = c + O_P(p^{-1/2})$.

As for $B_2$, it can be shown that  $E(B_2)=0$ and
$$\textrm{Var}(B_2)=\frac{l^2(l-1)^2}{p^4(l-m)^2}=o\left(p^{-2}\right) \enskip ,$$
from which it follows, by Chebyshev's inequality, that $B_2=O_P(p^{-1}).$

Combining all of the results above, under the white noise model, i.e., when
$\sigma_1=\sigma_2=\ldots=\sigma_l=1$,  we have $B = c+O_P(p^{-1/2})$ .
In the case that the spiked covariance model instead holds, i.e., when
$\sigma_1>\sigma_2\ldots>\sigma_m>\sigma_{m+1}=\ldots=\sigma_l=1$,
it can be shown that the impact on equation~\ref{B3} is to introduce
an additional term of $o_P(p^{-1})$.  The effect is therefore negligible on the final
result stated in the theorem, which involves an $O_P(p^{-1/2})$ term.
Intuitively, the value of the first $m$ eigenvalues $\sigma_i$ will not
influence the asymptotic behavior of the infinite sum in~\ref{B3}, which
is term B.

\subsection{Proof of Theorem 3}

Through a coordinate transformation, from $X$ to $V^TX$, we can, without loss
of generality, restrict our attention to the case where
$X\sim N(\mu,\Sigma)$,
for $\Sigma = \hbox{diag}(\sigma_1,\ldots,\sigma_l)$,
and our testing problem is of the form
$$H_0:\mu=0 \enskip vs \enskip H_1:\mu=(\underbrace{0,\ldots,0}_{d>k},\gamma,0,\ldots,0)^T$$
for some $\gamma>0$.  In other words, we test whether the underlying mean
is zero or differs from zero in a single component by some value $\gamma > 0$.

Consider first the expression for $\hbox{\sc Power}_Q$ in
(\ref{eq:Q.power.def}).  Under the above model,
$$Q = \sum_{j=k+1}^{l} Z_j^2\enskip ,$$
where the $Z_j$ are independent $N(\mu_j,1)$ random variables.
So, under the alternative hypothesis,
the sum of their squares is a non-central chi-square random variable,
on $l -k$ degrees of freedom, with non-centrality parameter
$||(\mu_{k+1},\ldots,\mu_{l})^T||^2_2= \gamma^2$.
We have by standard formulas that
$$E[Q] = (l-k) + \gamma^2$$
and
$$\hbox{Var}(Q) = 2(l-k) + 4\gamma^2\enskip .$$
Using these expressions and the normal-based expression for
power defining (\ref{eq:Q.power.def}), we find that
$$\hbox{\sc Power}_Q(\gamma) =
   \mathbb{P}\left(Z \ge z_{1-\alpha}\sqrt{\frac{(l-k)}{(l-k)+2\gamma^2}}
            - \frac{\gamma^2}{\sqrt{2(l-k) + 4\gamma^2}}\right) \enskip ,$$
as claimed.

Now consider the expression for $\hbox{\sc Power}_{Q^*}$ in
(\ref{eq:Qstar.power.def}), where we write $Q^\ast= Y^T M Y$
for $Y = \frac{1}{\sqrt{p}}\Phi^T X$ and $M=(I-U_k {U_k}^T)$.
Under the null hypothesis, we have
$$E_{X|\Phi}(Q^*) = tr\left(M \, \frac{1}{p} \Phi^T\Sigma \Phi\right)$$
and
$$\hbox{Var}_{X|\Phi}(Q^*) = 2tr\left(M\frac{1}{p}\Phi^T\Sigma\Phi M\frac{1}{p}\Phi^T\Sigma\Phi\right) \enskip .$$
Call these expressions $\epsilon$ and $\nu$, respectively.  Under the
alternative hypothesis, the same quantities take the form
$$E_{X|\Phi}(Q^*) = \epsilon +\gamma^2 \tilde B \enskip$$
and
$$\hbox{Var}_{X|\Phi}(Q^*) = \nu +4\gamma^2 \tilde A \enskip ,$$
respectively, where
\begin{equation}
\gamma^2 \tilde A =
\frac{1}{p} \mu^T\Phi [ M (\frac{1}{p}\Phi^T \Sigma\Phi) M]\Phi^T\mu
\label{eq:Ascaling}
\end{equation}
and
\begin{equation}
\gamma^2 \tilde B = \frac{1}{p} \mu^T\Phi M \Phi^T\mu \enskip .
\label{eq:Bshift}
\end{equation}
Arguing as above, we find that
$$\hbox{\sc Power}_{Q^*}(\gamma) = \mathbb{P}\left (Z>
   z_{1-\alpha}\sqrt{\frac{\nu}{\nu+4\gamma^2 \tilde A}}
     -\frac{\gamma^2 \tilde B}{\nu+4\gamma^2 \tilde A}
  \right)
  \enskip .$$

By Theorem~\ref{thm:Var.Of.Qstar},
we know that $\nu=2(l-k)(c+1+O_P(p^{-1/2}))$.
Ignoring the higher-order stochastic error term we therefore have
$$ \hbox{\sc Power}_{Q^*}(\gamma)  = P\left(Z \ge z_{1-\alpha}
        \sqrt{\frac{(c+1)(l-k)}{(c+1)(l-k)+2\gamma^2 \tilde A}}
            - \frac{\gamma^2 \tilde B}{\sqrt{2(c+1)(l-k) +
             4\gamma^2 \tilde A}} \right) \enskip .$$
This expression is the same as that in the statement of
Theorem~\ref{thm:power}, up to a re-scaling by a factor of $c+1$.
Hence it remains for us to show that
$$\tilde A=c+1+O_P(p^{-1/2}) \quad\hbox{and}\quad
  \tilde B=1+O_P(p^{-1/2}) \enskip .$$

Our problem  is simplified under transformation by the rotation $\Phi \to \Phi O$, where
$O_{p\times p}$ is an arbitrary orthonormal rotation matrix.  If we similarly apply
$$\Sigma^\ast\to O^T\Sigma^\ast O,\, U\to O^TU,\, \hbox{and}\, M\to O^TMO,$$
then $\tilde A$ and $\tilde B$ remain unchanged in (\ref{eq:Ascaling}) and (\ref{eq:Bshift}).
Recall that $\Sigma^\ast=U\Lambda^* U^T$, where $\Lambda^*=diag(s_1,\ldots,s_p)$,
and $\mu=(0,\ldots,0,\gamma,0,\ldots,0)^T$, with $\gamma$ in the $d+1> k$ location.
Choosing $O=U$ and denoting $\Phi U=(\eta_{ij})$, straightforward calculations yield that
$$\tilde A=\frac{1}{p}\sum_{j=k+1}^ps_j\eta_{dj}^2$$
and
$$\tilde B=\frac{1}{p}\sum_{j=k+1}^p\eta_{dj}^2 \enskip .$$

Now write $\Phi^T = [\Phi_k^T, \Psi_k^T]^T$, where $\Phi_k$ denotes the first $k$ rows of $\Phi$, and
$\Psi_k$, the last $l-k$ rows.  The elements $\eta_{dj}$ in the two sums immediately above  lie in the $d$-th row
of the product of $\Psi_k$ and the last $p-k$ columns of $U$.  By Paul~\cite[Thm 4]{debashis}, we know that if
$\sigma_m$, the last leading eigenvalue in the spiked covariance model, is much greater than 1, and $l,p\to \infty$
 such that $\frac{l}{p}=c+o(p^{-1/2})$, then the distance between the subspaces $\textrm{span}\{\Phi_k\}$ and $\textrm{span}\{U_k\}$ diminishes to zero. Asymptotically, therefore, we may assume that these two subspace coincide.
Hence, since $\Psi_k$ is statistically independent of $\Phi_k$, it follows that $\Psi_k$ is asymptotically independent of
$U_k$, and therefore of the orthogonal complement of $U_k$, i.e., the last $(p-k)$ columns of $U$.  As a result,
the elements in $(\eta_{d,k+1},\ldots,\eta_{d,p})^T$ behave asymptotically like independent and identically
distributed standard normal random variables.
Applying  Chebyshev's inequality in this context, it can be shown that
$$\tilde A=c+1+O_P(p^{-1/2}) \quad\hbox{and}\quad \tilde B=1+O_P(p^{-1/2}).$$
Rescaling by $(c+1)$, the expressions for $A$ and $B$ in Theorem 4 are obtained.

\subsection{Proof of Theorem 4}
Let $M=I- U_k U_k^T$ and $\hat M = I - \hat{U}_k \hat{U}_k^T$.  If we use the sample covariance
$\widehat{\Sigma^{\ast}}=\frac{1}{n}({\bf Y}-\bar{Y})({\bf Y}-\bar{Y})^T$ to
estimate $\Sigma^{\ast}$, we will observe the residual
$\widehat{Q}^{\ast}=Y^T\hat{M}Y$ instead of $Q^{\ast}=Y^T MY$.  To prove the theorem it is sufficient 
to derive expressions for $E_{X|\Phi}( \hat{Q}^*)$ and $\hbox{Var}_{X|\Phi}(\hat{Q}^*)$ under 
the null and alternative hypothesis in (\ref{eq:hypotheses}), as these expressions  are what inform the components of the critical value in the power calculation.  
Our method of proof involves re-expressing  $E_{X|\Phi}( \hat{Q}^*)$ and $\hbox{Var}_{X|\Phi}(\hat{Q}^*)$ in terms of $M$ and $\hat M - M$ and
showing that those terms involving the latter are no larger than the error terms associated with the former in Theorems~1, 2, and~3.

Begin by considering the mean and writing
$$E_{X|\Phi}( \hat{Q}^*) = E_{X|\Phi}( Q^*) + E_{X|\Phi}(\hat Q^* - Q^*) \enskip .$$
We need to control the term
\begin{eqnarray}
E_{X|\Phi}(\hat Q^* - Q^*) & = & E_{X|\Phi}[ Y^T(\hat M - M)Y] \nonumber \\
                                              & = & tr\left[ (\hat M - M)\Sigma^*\right] + \frac{1}{p} \mu^T \Phi (\hat M - M)\Phi^T \mu \enskip .
\label{eq:exp.error.term}
\end{eqnarray}
Under the null hypothesis the second term in (\ref{eq:exp.error.term}) is zero, and so to prove (\ref{eq:Exp.Of.Qhatstar})
we need to show that the first term is $O_P(1)$.

Note that, without loss of generality, we may write
$\Sigma^*= \Sigma^*_1 + \Sigma^*_2$, where $\Sigma^*_1 = (1/p) \tilde \Phi^T_1 \Lambda_1 \tilde\Phi_1$
and $\Sigma^*_2 = (1/p) \tilde \Phi^T_2 \tilde\Phi_2$, for $\tilde \Phi^T = [\tilde\Phi_1^T, \tilde \Phi_2^T]$ a random matrix
of independent and identically distributed standard Gaussian random variables and $\Lambda_1 = \hbox{diag}(\sigma_1,\ldots,\sigma_m)$.
Then using~\cite[Thm II.1]{park1996trace}, with $D=-\Sigma^*_2$ in the notation of that paper, it follows that
\begin{equation}
\left\vert tr\left[ (\hat M - M)\Sigma^*\right] \right\vert \le
\max \left( |\lambda_1(\hat M - M)|, |\lambda_p(\hat M - M)|\right)\,
	\left[ tr(\Sigma^*) - tr(\Sigma^*_2) \right] + tr\left[ (\hat M - M)\Sigma^*_2\right] \enskip ,
\label{eq:big.trace.bnd}
\end{equation}
where we use $\lambda_i(\cdot)$ generically here and below to denote the $i$-th largest eigenvalue of its argument.

For the second term in the right-hand side of (\ref{eq:big.trace.bnd}), write $\hat M - M = U_kU_k^T - \hat U_k \hat U_k^T$.
Using a result attributed to Mori (appearing as Lemma~I.1 in~\cite{park1996trace}), we can write
$$\lambda_p(\Sigma^*_2)tr(U_k U_k^T) \le tr( U_k U_k^T\, \Sigma_2^*) \le \lambda_1(\Sigma^*_2) tr(U_kU_k^T) \enskip ,$$
 and similarly for $\hat U_k\hat U_k^T$ 
in place of $U_kU_k^T$. Exploiting the linearity of the trace operation and the fact that $\hbox{rank}(\hat U_k \hat U_k^T) = \hbox{rank}(U_kU_k^T) = k$, we
can bound the term of interest as
$$\left\vert tr[ (\hat M - M)\Sigma^*_2] \right\vert \le k[ \lambda_1(\Sigma^*_2) - \lambda_p(\Sigma^*_2)]\enskip .$$
  However, $\lambda_1$ and
$\lambda_p$ are equal to $c+o(p^{-1/2})$ times the largest and smallest eigenvalues of a sample covariance of standard Gaussian random variables, 
the latter which converge almost surely to the right and left endpoints of the Marchenko-Pastur distribution (e.g., \cite{bai}), which in this setting take the values
$[1+(1/c)^{1/2}]^2$ and $[1-(1/c)^{1/2}]^2$, respectively.  Hence, $tr[(\hat M - M)\Sigma^*_2] = O_P(1)$.

Now consider the factor $tr(\Sigma^*) - tr(\Sigma^*_2)$ in the first term of the right-hand side of (\ref{eq:big.trace.bnd}).  We have shown that
$tr(\Sigma^*) = tr(\Sigma) + O_P(1)$.  At the same time, we note that $tr(\Sigma^*_2) = l( 1 + O_P((pl)^{-1/2})$, being proportional to the normalized trace of a matrix whose entries are independent and identically distributed copies of averages of $l-m$ independent and identically distributed chi-square random variables on one degree
of freedom.  Therefore, and recalling the spiked covariance model, we find that 
$tr(\Sigma^*) - tr(\Sigma^*_2) = \sum_{i=1}^m (\sigma_i - 1) + O_P(1)$.

At the same time, the factor multiplying this term, i.e., the largest absolute eigenvalue of $\hat M - M$,
 is just the operator norm $||\hat M - M||_2$ and hence bounded above by 
the Frobenius norm, $||\hat M - M||_F$.  We introduce the notation $P_j$ for the $j$-th column of
$U$ times its transpose, and similarly, $\widehat P_j$, in the case of $\hat U$.
Then $\hat M - M = \sum_{j=1}^k (P_j - \widehat P_j)$ and 
$$||\hat M - M||_F \le \sum_{j=1}^k \left\| \widehat P_j - P_j\right\|_F \enskip .$$
To bound this, we use a result in Watson~\cite[App B, (3.8)]{watson}, relying on a
multivariate central limit theorem,
$$n\|\widehat{P_j}-P_j\|_F^2\to2\sum_{k\neq
j}\frac{tr(P_jGP_kG)}{(s_j-s_k)^2} \enskip $$
in distribution, as $n\rightarrow \infty$, where $G$ is a random
matrix whose distribution depends only on $\Sigma^*$ and recall
$(s_1,\ldots,s_p)$ are the eigenvalues of $\Sigma^*$.  So 
$||\hat M - M||_2 = O_P(n^{-1/2})$.  

Therefore, the left-hand side of (\ref{eq:big.trace.bnd}) is $O_P(1)$ and (\ref{eq:Exp.Of.Qhatstar}) is established.
Now consider the second term in (\ref{eq:exp.error.term}), which must be controled under the alternative hypothesis.
This is easily done, as we may write
$$\left\vert \frac{1}{p} \mu^T \Phi (\hat M - M)\Phi^T \mu \right\vert 
\le \frac{1}{p} || \Phi^T \mu||^2_2 \,\, ||\hat M - M||_2 \enskip ,$$
and note that the first term is $O_P(1)$ while the second is $O_P(n^{-1/2})$.  Therefore, under the assumption that $n\ge p$,
the entire term is $O_P(p^{-1/2})$, which is the same order of error to which we approximate $\tilde B$ in (\ref{eq:Bshift}) 
in the proof of Theorem~3.  Hence, the contribution of the mean to the critical value in (\ref{eq:Qstarhat.power.eqn}), using $\hat\Sigma^*$,  is the
same as in (\ref{eq:Qstar.power.eqn}), using $\Sigma^*$.

This completes our treatment of the mean.  The variance can be treated similarly, writing
$$\hbox{Var}_{X|\Phi}(\hat Q^*) = \hbox{Var}_{X|\Phi}(Q^*) + \hbox{Var}_{X|\Phi}(\hat Q^* - Q^*)
						+ 2 \hbox{Cov}_{X|\Phi}(Q^*, \hat Q^* - Q^*) $$
and controling the last two terms.  The first of these two terms takes the form
\begin{equation}
\hbox{Var}_{X|\Phi}(\hat Q^* - Q^*) = 2 tr\left[ (\hat M - M)\Sigma^*\right]^2 +
                                                                  \frac{4}{p}\mu^T \Phi \left[ (\hat M - M) \Sigma^* (\hat M - M)\right] \Phi^T \mu \enskip ,
\label{eq:var.error}
\end{equation}
and the second,
\begin{equation}
\hbox{Cov}_{X|\Phi}(Q^*, \hat Q^* - Q^*) = 2 tr\left[ M \Sigma^* (\hat M - M)\Sigma^*\right] +
                                                                  \frac{4}{p}\mu^T \Phi \left[  M \Sigma^* (\hat M - M)\right] \Phi^T \mu \enskip .
\label{eq:cov.error}
\end{equation}

Again, under the null hypothesis, the second terms in (\ref{eq:var.error}) and (\ref{eq:cov.error}) are zero.  Hence, to establish
(\ref{eq:Var.Of.Qhatstar}), it is sufficient to show that the first terms in (\ref{eq:var.error}) and (\ref{eq:cov.error}) are $O_P(p^{1/2})$.
We begin by noting that
$$tr\left[ (\hat M - M)\Sigma^*\right]^2 \le tr\left[ (\hat M - M)^2 (\Sigma^*)^2 \right]
                  \le tr\left[ (\hat M - M)^2\right] \, tr\left[ (\Sigma^*)^2\right] \enskip ,$$
where the first inequality follows from~\cite[Thm 1]{chang1999matrix}, and the second, from Cauchy-Schwartz.
Straightforward manipulations, along with use of~\cite[Lemma I.1]{park1996trace}, yields that
$tr(\hat M-M)^2 \le 2k ||\hat M - M||_2 = O_P(n^{-1/2})$.  At the same time, we have that
$$tr(\Sigma^*)^2  \le  \lambda_1(\Sigma^*) \, tr(\Sigma^*) 
                           =   \left[ \lambda_1(\Sigma) + O_P(p^{-1/2})\right]\, \left[tr(\Sigma) + O_P(1)\right] = O_P(l)\enskip .$$
Therefore, under the assumptions that $n\ge p$ and $l/p = c + o(p^{-1/2})$, we are able to control the relevant error term in (\ref{eq:var.error})
as $O_P(n^{-1/2})O_P( l) = O_P(p^{1/2})$.  

Similarly, using~\cite[Lemma I.1]{park1996trace} again, we have the bound
$$\left\vert tr\left[ M \Sigma^* (\hat M - M)\Sigma^*\right] \right\vert \le ||(\hat M - M)\Sigma^* ||_2 \,\, tr(M\Sigma^*) \enskip .$$
The first term in this bound is $O_P(n^{-1/2})$, while the second is $O_P(l)$, which allows us to control the relevant error term in (\ref{eq:cov.error})
as $O_P(p^{1/2})$.  As a result, under the null hypothesis, we have that
$\hbox{Var}_{X|\Phi}(\hat Q^*) = \hbox{Var}_{X|\Phi}(Q^*) + O_P(p^{1/2})$, which is sufficient to establish (\ref{eq:Var.Of.Qhatstar}),
since $\hbox{Var}_X(Q) = O(l) = O(p)$.

Finally, we consider the second terms in (\ref{eq:var.error}) and (\ref{eq:cov.error}), which must be controled as well under the alternative
hypothesis.  Writing
$$\mu^T \Phi \left[ (\hat M - M) \Sigma^* (\hat M - M)\right] \Phi^T \mu \le ||\Phi^T\mu||^2_2 \, ||\hat M - M||^2_2 \, ||\Sigma^*||_2$$
and
$$\left\vert \mu^T \Phi \left[  M \Sigma^* (\hat M - M)\right] \Phi^T \mu \right\vert \le 
     ||\Phi^T\mu||^2_2 \, ||\hat M - M||_2 \, ||\Sigma^*||_2 \, ||M||_2 \enskip ,$$
it can be seen that we can bound the first of these expressions by $O_P(1)$, and the second, by $O_P(p^{1/2})$.  Therefore, the combined contribution of
the second terms in  (\ref{eq:var.error}) and (\ref{eq:cov.error}) is $O_P(p^{-1/2})$, which is the same order to which we approximate $\tilde A$
in (\ref{eq:Ascaling}) in the proof of Theorem~3.  Hence, the contribution of the variance to the critical value in (\ref{eq:Qstarhat.power.eqn}), using
$\hat\Sigma^*$, is the same as in (\ref{eq:Qstar.power.eqn}), using $\Sigma^*$.

\bibliographystyle{plain}
\bibliography{comp-pca-subspace-biblio}

\end{document}